# Causal analysis of competing atomistic mechanisms in ferroelectric materials from high-resolution Scanning Transmission Electron Microscopy data

Maxim Ziatdinov,[1,5] Chris Nelson,[1] Xiaohang Zhang,[2] Rama Vasudevan,[1] Eugene Eliseev,[3] Anna N. Morozovska,[4] Ichiro Takeuchi,[2] and Sergei V. Kalinin[1,a]

[1] The Center for Nanophase Materials Sciences, Oak Ridge National Laboratory, Oak Ridge, TN 37831, USA

[2] Department of Materials Science and Engineering, University of Maryland, College Park, MD 20742, USA

[3] Institute for Problems of Materials Science, National Academy of Sciences of Ukraine, Krjijanovskogo 3, 03142 Kyiv, Ukraine

[4] Institute of Physics, National Academy of Sciences of Ukraine, 46, pr. Nauky, 03028 Kyiv, Ukraine

[5] Computational Sciences and Engineering Division, Oak Ridge National Laboratory, Oak Ridge, TN 37831, USA

Machine learning has emerged as a powerful tool for the analysis of mesoscopic and atomically resolved images and spectroscopy in electron and scanning probe microscopy, with the applications ranging from feature extraction to information compression and elucidation of relevant order parameters to inversion of imaging data to reconstruct structural models. However, the fundamental limitation of machine learning methods is their correlative nature, leading to extreme susceptibility to confounding factors. Here, we implement the workflow for causal analysis of structural scanning transmission electron microscopy (STEM) data and explore the interplay between physical and chemical effects in ferroelectric perovskite across the ferroelectric-antiferroelectric phase transitions. The combinatorial library of the Sm – doped $BiFeO_3$ is grown to cover the composition range from pure ferroelectric BFO to orthorhombic 20% Sm-doped BFO. Atomically resolved STEM images are acquired for selected compositions and are used to create a set of local compositional, structural, and polarization field descriptors. The information-geometric causal inference (IGCI) and additive noise model (ANM) analysis are used to establish the pairwise causal directions between the descriptors, ordering the data set in the causal direction. The causal chain for IGCI and ANM across the composition is compared and suggests the presence of common causal mechanisms across the composition series. Ultimately, we believe that the causal analysis of the multimodal data will allow exploring the causal links between multiple competing mechanisms that control the emergence of unique functionalities of morphotropic materials and ferroelectric relaxors.

[a] sergei2@ornl.gov

Functionality of material systems such as morphotropic phase boundary systems,[1-4] ferroelectric relaxors,[5-8] spin and cluster glasses,[9-12] charge ordered manganites,[13-17] are determined by the complex interplay between structural, orbital, chemical, spin and other degrees of freedom.[18,19] Traditionally, these materials system has been explored via the combination of macroscopic physical property measurements and scattering techniques, with the theoretical counterpart being provided via combination of analytical and numerical methods. Given that the physics of these materials is ultimately linked to the emergence of frustrated degenerate ground states driven by competing interactions, analyses based on the macroscopically averaged descriptors such as concentrations, order parameter fields, etc. provide only limited insight into the generative and especially causal physics of these materials systems.

Progress in the high-resolution imaging techniques have allowed visualization of these materials systems to the atomic level. Techniques such as Scanning Tunneling Microscopy (STM) have provided insight into electronic structure of surfaces and superconductive and magnetic order parameters.[20,21] Scanning Transmission Electron Microscopy (STEM) enabled studies of chemical composition down to the single atom level[22-24] and, via quantitative mapping of structural distortions, enabled visualization of order parameter fields such as polarization[25-28], tilts[29-31], and mechanical[25,32-35] and chemical [35-37] strains. However, this emergence of data brings the challenge of analysis of systems with multiple spatially distributed degrees of freedom, including determination of both the functional laws connecting the functionalities and structure and the causal links that define the cause and effect relationship in the non-stationary and non-ergodic systems.

Recently, machine learning (ML) has emerged as a powerful tool for the analysis of mesoscopic and atomically resolved images and spectroscopy in electron and scanning probe microscopy.[38-41] The applications ranging from feature extraction[42] to information compression and elucidation of relevant order parameters[43] to inversion of imaging data to reconstruct structural models have been demonstrated. However, the fundamental limitation of the vast majority of machine learning methods is their correlative nature, leading to extreme susceptibility to confounding factors and observational biases.[44,45] While in classical statistical methods methodologies to address confounder- or selective bias induced phenomena such as Simpson paradox are established,[46] the complex and often non-transparent nature of modern machine learning tools such as deep neural networks renders them extremely prone to misinterpretation.

We pose that correlative machine learning provides a reliable and powerful tool in cases when the causal links are well established, as is atom finding in SPM and STEM and analysis of 4D STEM data when this condition is satisfied. Notably, ML applications in theory generally fall under this category since the causal links are postulated. Alternatively, ML methods work well when the confounding factors are effectively frozen via the narrowness of experimental conditions or experimental system. However, both these conditions are violated for experimental studies, when causal relationships are known only partially (and are in fact often the target of study) and confounding and observational bias factors (composition uncertainty, microscope tuning, contaminations) are abundant.

One approach to explore the generative physical models from the microscopic data is based on the fit to the relevant mesoscopic or atomistic models, i.e. discovering the generative physical models. On the mesoscopic level, direct match between the solution between Ginzburg-Landau equations and order parameter fields determined from the atomically resolved data can be used to determine the interface terms via corresponding boundary conditions,[47] as well as the nature of coupling and gradient terms.[39] Statistical distance minimization can be used to directly match the discrete data to lattice model, e.g. to reconstruct the interaction parameters.[48-51] However, even when the functional laws describing the system are known, this level of the description is not sufficient to establish the **causal** mechanisms active in the system. As a simple example, the knowledge of the ideal gas law does not establish whether pressure is cause or an effect of the volume change unless the character of the process is established.

In many cases, it can be argued that the causal effects can be estimated based on the energy scales of corresponding phenomena, e.g. magnetic properties driven by relatively weak energy scales are unlikely to affect atomic structure. However, this is not the case when the energy scales are comparable, or when depolarization and global effects become significant. For example, while the magnetization energy density per volume can be small, concentration of magnetically induced mechanical stresses can lead to stress corrosion at the domain walls. Specifically, for ferroelectric materials it is generally assumed that cationic order is frozen at the state of material formation, and then polarization field evolves to accommodate average polarization instability and local pinning. However, it is known that ions can redistribute to compensate polarization, with examples including segregation at the domain walls, memory effects, etc.[2,52] Hence, for non-equilibrium and non-ergodic materials the question of cause and effect become paramount. For example, does polarization align to the cationic disorder or does polarization instability at the morphotropic phase boundaries drive the cationic disorder?

More generally, being able to answer causal questions is required both for meaningful applications of machine learning techniques and inferring the materials physics, since causal knowledge allows avoiding correlative, but incorrect conclusions, explore counterfactuals and interventions,[53] i.e. realistic strategies for materials development. Correspondingly, we argue that analyzing causal relationships from the observations of atomically resolved degrees of freedom is key for understanding the physics of non-ergodic systems.

Here, we implement the workflow for causal analysis of STEM data and explore the interplay between physical and chemical effects in ferroelectric perovskite across the ferroelectric-antiferroelectric phase transitions. The combinatorial library of the Sm – doped $BiFeO_3$ is grown to cover the composition range from pure ferroelectric BFO to orthorhombic 20% Sm-doped BFO.[54-57] Atomically resolved STEM images are acquired for selected compositions and are used to create a set of local compositional, structural, and polarization field descriptors. The information-geometric causal inference (IGCI) and the additive noise model (ANM) are used to establish the pairwise causal directions between the descriptors, ordering the data set in the causal direction. The causal chain for IGCI and ANM across the composition is compared, suggesting the similarity of causal mechanisms across the Sm-BFO compositions.

The combinatorial library of $Bi_{1-x}Sm_xFeO_3$ ($0 \leq x \leq 0.2$) was fabricated on a $SrTiO_3$ (001) substrate after the deposition of a $SrRuO_3$ layer (30 nm). The chemical compositions at different positions across the substrate were characterized by wavelength dispersive spectroscopy (WDS) measurements. X-ray diffraction results (Fig. 1) indicate that the (002) peak of the $Bi_{1-x}Sm_xFeO_3$ layer gradually moves towards a higher angle as the Sm doping concentration ($x$) increases. As shown in Fig. 1 B-D, the representative piezoresponse force microscopy (PFM) images indicate that the domain structure in the combinatorial library changes from strip domains in the pure $BiFeO_3$ side to mosaic-like domains at an intermediate doping level (x ≈ 0.08), and eventually no domain structure can be identified for the highest doping level (x ≈ 0.2).

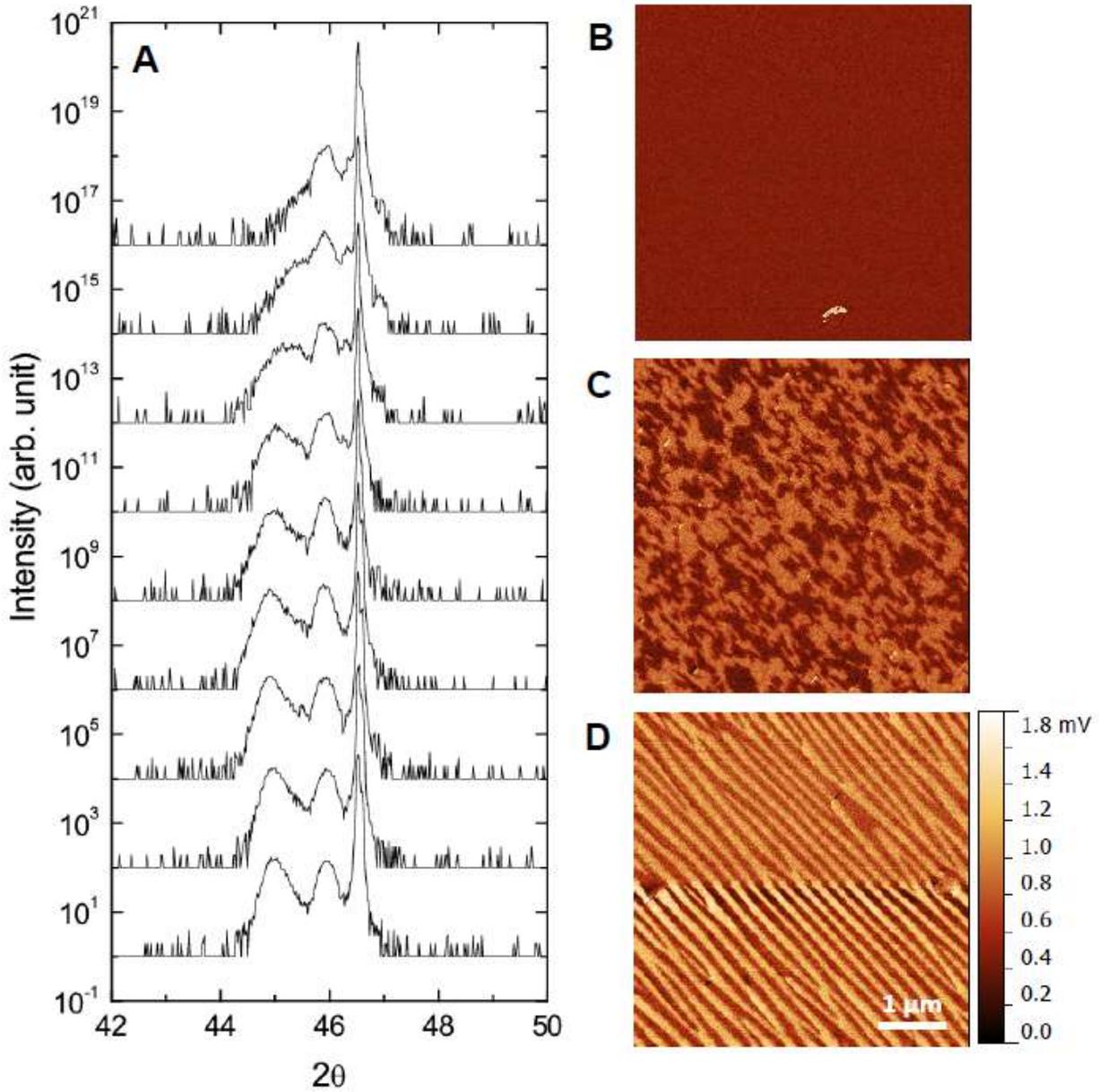

**Figure 1.** (A) X-ray diffraction results obtained from a Bi$_{1-x}$Sm$_x$FeO$_3$(120 nm)/SrRuO$_3$(30 nm) sample fabricated on a (001) SrTiO$_3$ substrate, where $0 \leq x \leq 0.2$. The results are equally spaced for clarity with the bottom curve corresponding to BiFeO$_3$ and the top curve corresponding to Bi$_{0.8}$Sm$_{0.2}$FeO$_3$. (B)-(D) representative piezoresponse force microscopy (PFM) images for a highly doped region ($x \approx 0.2$), an intermediately doped region ($x \approx 0.08$), and an undoped region ($x \approx 0$), respectively.

The TEM samples were prepared for three sites along the gradient composition sample with nominal compositions of 0%, 7% and 20% Sm doping (see methods). STEM data was collected from the [100] pseudocubic zone axis using High-Angle Annular Dark Field (HAADF) detector imaging as shown in Fig. 2, thereby providing the projected atomic structure as well as compositional information by the atomic column intensity (which scales by $\sim Z^2$). As the concentration of Sm increases through the sample series there is an observed phase transition from the prototypical rhombohedral ferroelectric phase [58] of BiFeO$_3$ (Fig. 2A) to an antiferrodistortive orthorhombic phase[59] at 20% Sm (Fig. 2C). The transition is readily observable in maps of the polar atomic displacement between the A-site and B-site cation sublattices, ***P***, which is shown for the three compositions in Figure 2.

For the pure BiFeO$_3$ phase ***P*** is a proxy for the electrical dipole moment and the distribution in Fig. 2A illustrates the polydomain structure characteristic of an *r*-phase ferroelectric including a 109° (vertical) and 180° (inclined) domain walls. The distribution of ***P*** in the 20% Sm composition depicts the large oscillation of $P_y$ corresponding to the antiferrodistortive orthorhombic structure (Fig. 2C). The intermediate 7% Sm composition exhibits a mixed structure, with the small domains identifiable to both structures appear in the near interface region (Fig. 1B).

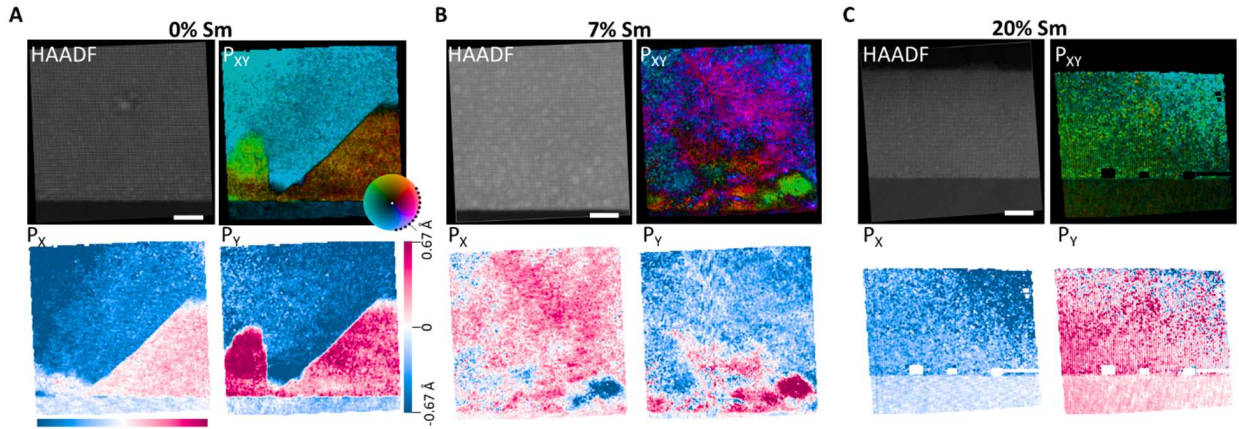

**Figure 2. Atomic resolution HAADF STEM imaging of Sm$_x$:Bi$_{1-x}$FeO$_3$.** Datasets for 0% Sm **(A)** 7% Sm **(B)** and 20% Sm **(C)** are shown. For each, the top left panel is the HAADF image, top right is the corresponding map of the polar displacement vector ***P***, bottom left and right are the x- and y- axis components of ***P***. Color scales are identical across the compositions according the legends in **(A)**. Scalebars are 10nm. ***P*** vectors and components have scale limits of ±0.67Å.

To describe the local materials behaviors, we parametrize the data choosing the perovskite unit cell as the basis. We introduce a set of descriptors for the local material behavior based on the properties and distribution of the 5 cation atomic columns in each unit cell. These are outlined in Table 1, along with the corresponding calculations. Unit-cell parameters are defined from local neighborhood atomic columns of HAADF STEM data corresponding to a five cation perovskite-type cell: corner A-sites $A_1$, $A_2$, $A_3$, $A_4$ and central B-site $B_1$ (labels in Fig 2A). Parameters include structural descriptors from positional data regarding unit cell size and shape (*a*, *b*, *a/b*, *θ*, *Vol*), compositional information from atomic HAADF intensity & distribution ($I_1$, $I_2$, $I_3$, $I_4$, $I_5$), and electrical polarization information from non-centrosymmetric displacement of the A- and B-site sublattices (***P***). Examples for several of these descriptors for a HAADF STEM unit cell are illustrated in Figure 2A. We here define *a* and *b* as the two lattice vectors of the unit cell connecting A-site corner positions, their internal angle *θ*, magnitude ratio *a/b*, and total cell volume *Vol*. $I_1$ corresponds to the mean atomic column intensity and scales with the sample mass-thickness. $I_2$-$I_5$ correspond to internal asymmetries. Notably, $I_5$ corresponds to the intensity ratio between cation sublattices, thus readily distinguishes the $Sm_xBi_{1-x}FeO_3$ film and $SrTiO_3$ substrate. Our choice of basis also includes several internal gradient terms including A-site intensity asymmetries in $I_2$-$I_4$ and the gradient of the *a* and *b* vector between opposed edges of the unit cell (denoted as ***ab*Δ**). Moreover, additional gradient terms can be derived using a larger basis or calculated across multiple neighbor cells.

**Table 1 – Unit cell descriptors**

| Sym. | U.C. Parameter Descriptions | Calculation |
|---|---|---|
| $\vec{a}$ | In-plane lattice vector | $\dfrac{(A_{2,xy} - A_{1,xy}) + (A_{4,xy} - A_{3,xy})}{2}$ |
| $\vec{b}$ | Out-of-plane lattice vector | $\dfrac{(A_{4,xy} - A_{1,xy}) + (A_{3,xy} - A_{2,xy})}{2}$ |
| $\overrightarrow{ab\Delta}$ | *ab* delta vector | $\dfrac{-(A_{2,xy} - A_{1,xy}) + (A_{4,xy} - A_{3,xy})}{2}$ |
| *θ* | Internal angle | $atan(\vec{a}\vec{b})$ |
| *a/b* | Tetragonality | $\dfrac{|a|}{|b|}$ |
| *Vol* | Internal volume | Volume of convex hull $A_{1,xy}$, $A_{2,xy}$, $A_{3,xy}$, and $A_{4,xy}$ |
| $I_1$ | Mean atomic HAADF intensity | $\dfrac{A_1 + A_2 + A_3 + A_4 + B_1}{5}$ |
| $I_2$ | A-site Asymmetry $A_1, A_2$ vs. $A_3, A_4$ | $\frac{(A_1 + A_2) - (A_3 + A_4)}{2}$ |
| $I_3$ | Asymmetry $A_1, A_3$ vs. $A_2, A_4$ | $\frac{(A_1 + A_3) - (A_2 + A_4)}{2}$ |
| $I_4$ | Asymmetry $A_1, A_4$ vs. $A_2, A_3$ | $\frac{(A_1 + A_4) - (A_2 + A_3)}{2}$ |
| $I_5$ | Asymmetry A-site vs. B-site | $\frac{(A_1 + A_2 + A_3 + A_4)}{4} - B_1$ |
| $\vec{P}$ | Polar displacement vector | $\frac{A_{1,xy} + A_{2,xy} + A_{3,xy} + A_{4,xy}}{4} - B_{1,xy}$ |

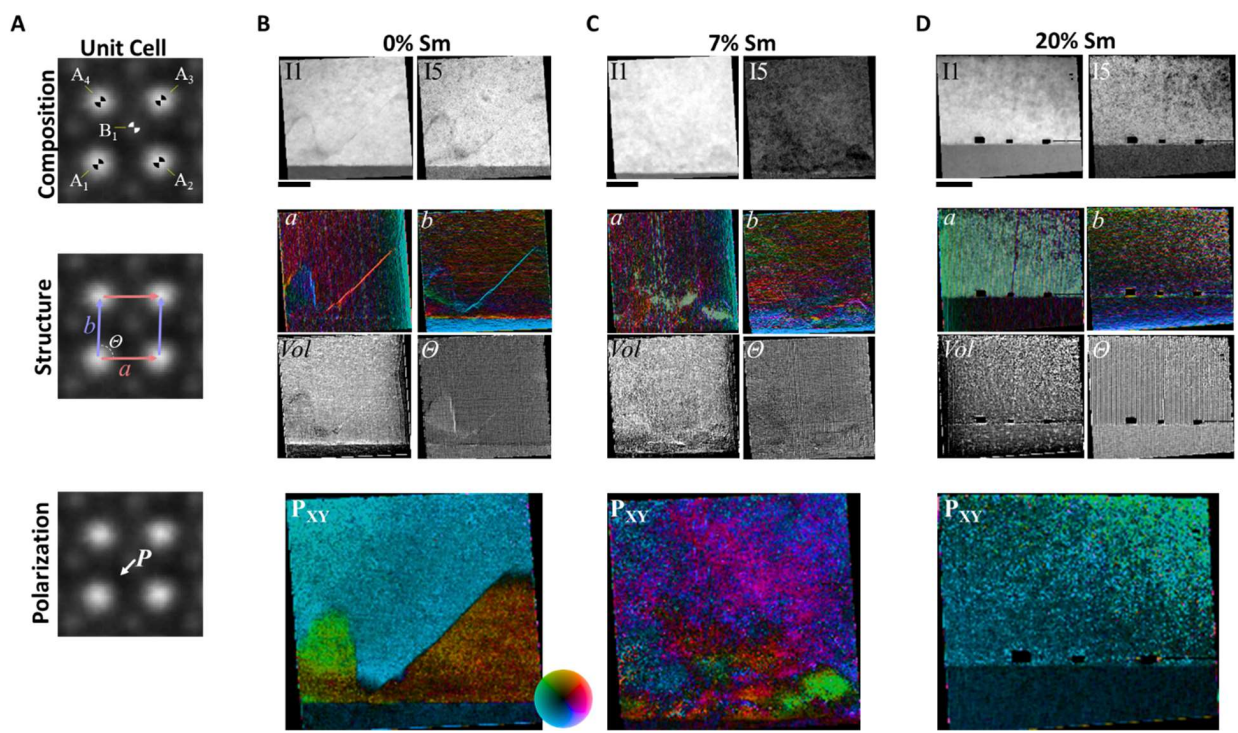

**Figure 3. Unit cell descriptor maps. A)** Descriptors are derived for each B-site centered 5-atom perovskite unit cell in the HAADF STEM images. **(B-D)** Selected descriptors are shown for the 0% **(B)** 7% **(C)** and 20% **(D)** compositions corresponding to local compositional ($I_1$, $I_5$), structural (*a*, *b*, *Vol*, *θ*), and polarization (*P*) information. Descriptors are mapped onto unit cell grid coordinates, each datapoint corresponds to 1 unit cell (~4Å). Spatial dimensions are not depicted at matched scales across the compositions (labeled scale bars below I1 plots are 40 unit cells) but data scales are matched (see Supplemental Materials Figs 1-3).

Real-space distributions of selected local physical and chemical descriptors of materials structure and functionality are shown in Figure 3 for the three compositions. The plots are depicted in unit-cell space, each data point corresponding to the local unit cell descriptor in an *a,b* addressed grid. The selected descriptor maps are categorized as compositional parameters (top) associated with the unit cell intensity represent a convolution of the (slowly changing) film thickness and local composition; structural parameters (mid) including lattice parameters, unit cell volume, and internal angle; and polarization components (bottom) describing the ferroelectric functionality. Using the unit-cell basis the alternating $P_y$ component from an atom-level mapping (Fig. 1C) is not observed here in Fig 2 D, but is instead captured by structural descriptors *a* and θ. Not depicted are internal descriptor or cross-unit cell gradient terms, distribution maps for internal gradient descriptors can be found in Supplemental Materials Figs 1-3.

To provide the physical context for causal analysis of the observables in the STEM experiment, we note that in general thermodynamics of ferroelectric materials can be described via Landau-Ginzburg-Devonshire (LGD) theory, where the energy of material can be represented as free energy functional

$$G = \int d^3 x \left( \Delta G_{AFD} + \Delta G_{FE} + \Delta G_{AFE} + \Delta G_{BQC} + \Delta G_{ST} + \Delta G_{EL} \right). \tag{1}$$

describe the antiferrodistortive (AFD), ferroelectric (FE), and antiferroelectric (AFE) long-range orders. AFD order is described by an axial vector, $\Phi_i$, that is perpendicular to the rotation plane of the oxygen octahedral tilts. FE and AFE long-range orders, which interact with AFD order, and transform to one another depending on the Sm content, are described by FE and AFE order parameters, $P_i = \frac{1}{2}(P_i^a + P_i^b)$ and $A_i = \frac{1}{2}(P_i^a - P_i^b)$, where $P_i^a$ and $P_i^b$ are the polarization components of two (or more) equivalent sublattices "*a*" and "*b*". Antiferromagnetic order is not included in Eq.(1), since its impact on AFD, FE and AFE orders are negligibly small, as a rule.

The individual AFD, FE, and AFE contributions are generally representable as the expansions in powers (2-4 for the second order, or 2-4-6 for the first order phase transitions) of corresponding order parameters, gradient terms defining the spatial behavior of the order parameter fields, coupling terms with the conjugate fields (electric, strain, strain gradient), and biquadratic coupling terms describing the interactions between order parameters. The important aspect of Ginzburg-Landau theory is that the free energy of material is generally non-local, since the order parameter and depolarizations fields can be found only from the solution of the boundary value problem. Therefore, in the most general description, the individual observables are linked through the integral transforms, representing extremely complex form of parameter coupling.

Here, we note that under some general conditions including macroscopic uniformity, this relationship can be simplified to yield the local non-linear relationship between state variables, with the non-local effects being represented via the unknown mean local fields. These nonlinear and non-local partial differential equations can be linearized around the specific ground state to give the linear relationship between the observed and non-observable parameters. Secondly, we note that in the presence of the strong composition fluctuations and nanodomains, the local fields will be the superposition of slowly varying (on the atomic level) depolarization fields and disorder related fields. Here, we aim to explore the causal relationships from the observed descriptor fields.

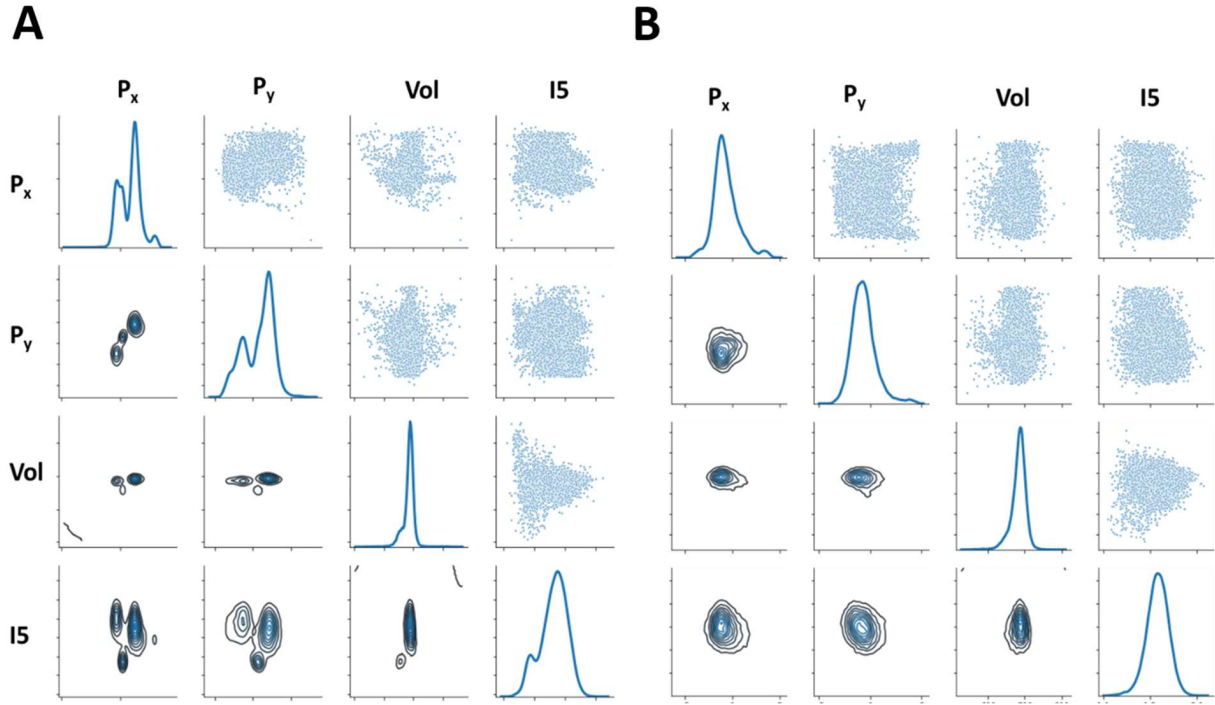

**Figure 4.** The correlation analysis of the STEM polarization, structural, and chemical descriptors for the (A) rhombohedral and (B) mixed phase (partial list).

The initial insight into the statistical properties of this material system can be deduced from the analysis of joint pairwise distributions as shown in Figure 4. Figure 4A shows the pairwise distributions for the set of parameters $P_x$, $P_y$, $V$, and $I_5$ for rhombohedral ferroelectric phase. Note that the full data set and analyses are contained in the accompanying Jupyter notebook (see

Methods section). Here, the diagonal elements contain the distribution functions for individual parameters. Notably both $P_x$ and $P_y$ distributions have 4 peaks, suggesting the more complex polarization distribution that can be expected in the system of ferroelectric domains (positive and negative) and substrate (0). In particular, $P_x$ component shows pronounced peak splitting for the non-zero polarization orientation. The distribution functions for molar volume $Vol$ and chemical variability $I_5$ show two clear peaks corresponding to ferroelectric materials and substrate respectively. Note that the width of the $I_5$ distribution is much broader, reflecting the higher variability or noise level in data.

The pairwise distributions between the descriptors are shown on the upper and lower triangular matrices. Here, the full data in the upper diagonal provides the general insight into the outliers. The kernel density estimates in the lower diagonal provide the insight into the statistically significant parts of the distributions. The pairwise distribution between $P_x$-$P_y$ has three peaks clearly corresponding to the two dominant domain orientations and the substrate. Similar structures are visible for $P_x$-$Vol$ and $P_y$-$Vol$ distributions, clearly showing the similar molar volumes for the ferroelectric phase and dissimilar molar volume for the substrate. Finally, the distribution function between $I_5$ and $P_x$, $P_y$, $Vol$ show complex multimodal distributions. It should be noted here that for ferroelectric phase the observations within the ferroelectric domain will impose the observational bias on the data; hence ideally the imaged volume should contain multiple domain or, alternatively, exceed the correlation length for observed variables.

Similar analysis for the Sm-rich phase 2 is shown in Fig. 4 B. In this case, the distribution functions peaks are clearly non-Gaussian, reflecting complex nature of the mixed phases. Similarly, pair distribution functions clearly show the asymmetry in the peak shapes, etc. Similar behavior is observed for the phase 3 (orthorhombic).

Even cursory examination of the distributions in Fig. 4 (or full versions available from the notebook) illustrates that these are generally not marginalizable, i.e. joint distributions between the parameters cannot be represented as the product of the marginal distribution functions. This in turn suggests the presence of the functional or causal link between the parameters. However, while some of these links can be speculated about (i.e. it can be argued that chemical fluctuations control order parameter distributions), multiple counter-examples such as cation redistribution during the aging of ferroelectric materials, etc. suggest that these "natural" explanations are not necessarily correct. Hence, we aim to analyze the causal distributions from the observational data.

Generally, analysis of causal relationships is one of the most complex problems in ML. For two observed variables, the number of possible causal relationships is limited and methodologies to establish directionality of causal link and presence of possible confounders are available. For more complex cases, the analysis of directed acyclic causal graphs has been explored by Pearl group.[44,46] However, analysis of the cause and effect relationships in the presence of cycles and feedbacks represents significantly more complex problem, and numerical schemes to address these have been reported only recently by Mooij and others.[60]

Here we explore two step approach for analysis of the possible causal relationships between the STEM observables. First, the causal directions are analyzed for all pairs of variables to yield pairwise causal relationships and represented as a "causal sieve" matrix. By construction, the matrix is antisymmetric with 1 and -1 elements. Secondly, the properties of graph which adjacency matrix is given by the "causal sieve" are explored.

To describe causal direction for two variables, we use and compare the information-geometric causal inference (IGCI)[61] and the additive noise model (ANM).[62] The IGCI method is based on the assumption of the independence of the 'cause' distribution and the conditional distribution of the 'effect' given the cause.[61,63,64] It can be shown, using an empirical slope-based estimator,[65] that $X$ causes $Y$ if

$$\sum_{j=1}^{N-1} \log \frac{|y_{j+1} - y_j|}{|x_{j+1} - x_j|} - \sum_{j=1}^{N-1} \log \frac{|x_{j+1} - x_j|}{|y_{j+1} - y_j|} < 0, \qquad (2)$$

and vice versa, where the $(x_j, y_j)$ pairs are ordered ascendingly according to $x$ in the first term and according to $y$ in the second term. IGCI was first assumed to be applicable only to noise-free observations where $Y = f(X)$ and $X = f^{-1}(Y)$ but was later shown (empirically) to work on noisy data as well.[64] The equation (2) was used for all the IGCI-based analysis of the cause-effect pairs in the current paper.

As a second pairwise causality check, we use the additive noise model (ANM) estimator for finding a causal direction from the observed data. The simple idea behind the ANM method is that the effect is a function of its cause plus a noise term independent of the cause.[66] In the ANM one performs the non-linear regression fitting, first for $X$ on $Y$ and then for $Y$ on $X$, and calculates the difference between test scores for the independence of residuals in both cases. The negative difference value implies that $X$ causes $Y$, while the positive value implies that $Y$ causes $X$.

For our analysis, we used a Gaussian process (GP) regressor with the squared exponential kernel. Because the exact inference of the GP regressor parameters is intractable for the datasets with ~$10^4$ points, we used the inducing points-based sparse GP approximation[67] with variational free energy (VFE) inference method, as implemented in Pyro's probabilistic programming language.[68] The inducing points were selected uniformly from the observation data points with a step of ~20. In addition to the GP regressor, we also added an option for choosing a 2-layer neural network as a regressor (see the accompanying Jupyter notebook). The independence of residuals test was done by calculating the Hilbert-Schmidt Independence Criterion (HSIC)[69] with the Gaussian kernel whose width was set to the median distance between points in input space.

The IGCI typically takes advantage of some specific features of the dataset, whereas the ANM tend to yield good results as long as the additive assumption holds.[70] Before applying to the experimental observations, both IGCI and ANM methods were first tested on the publicly available database of labeled cause-effect pairs[71] and the resultant accuracy in predictions (~64% and ~66%, respectively) was comparable to the results reported for the same database in the machine learning literature.[53] We then calculate a matrix of pairwise dependencies for the list of descriptors derived from the experimental descriptions.

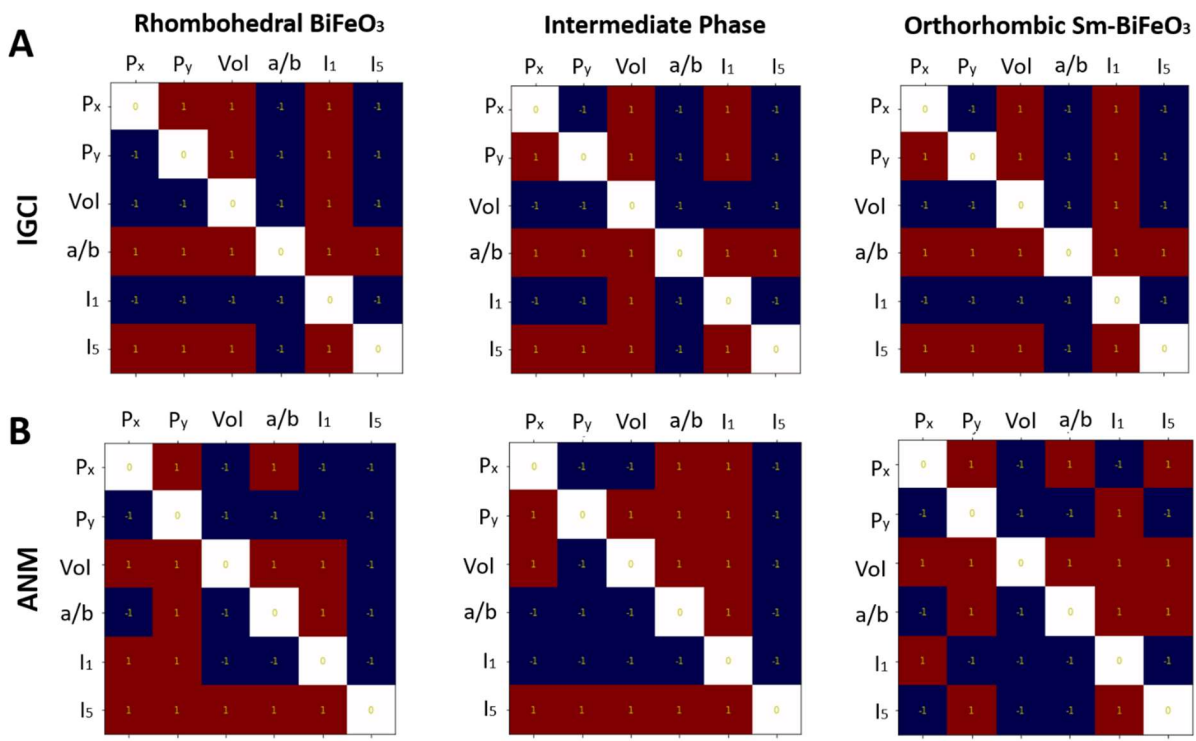

**Figure 5.** The causal sieve matrices for three Sm-BFO compositions. Shown are the results for the (A) IGCI and (B) ANM analyses.

The causal sieve matrices for three explored compositions are shown in Figure 5 for a selected subset of variables to enable ease of visualization. The analysis of the full set of structural, chemical, and polarization parameters is available in the accompanying notebook. Here, the positive 1 value means that the column value is identified as the cause, whereas the row variable is the effect, Col -> Row. Interestingly, the structures of the IGCI causal matrices in the studied cases are always such that for $N$ matrix one row contains $N$ positive entries, another row contains $N$-1 positive entries, etc. This implies that the descriptors can be formally ranked in the order of causal importance. Note that it does not imply that the system can be represented by linear causal chain. Rather, here we treat this behavior as observation.

Remarkably, the IGCI results are very similar across three different compositions. For rhombohedral and orthorhombic phases, the chemical composition parametrized as $I_1$ is identified as a cause variable affecting all other parameter but not affected by them. Second in importance is the molar volume *Vol*. For the intermediate phase, the *Vol* variable is higher in the "causal importance" than $I_1$. In all three cases, the ($I_1$, *Vol*) variables are followed by polarization components ($P_x$, $P_y$), differential chemical composition $I_5$, and finally by tetragonality $a/b$.

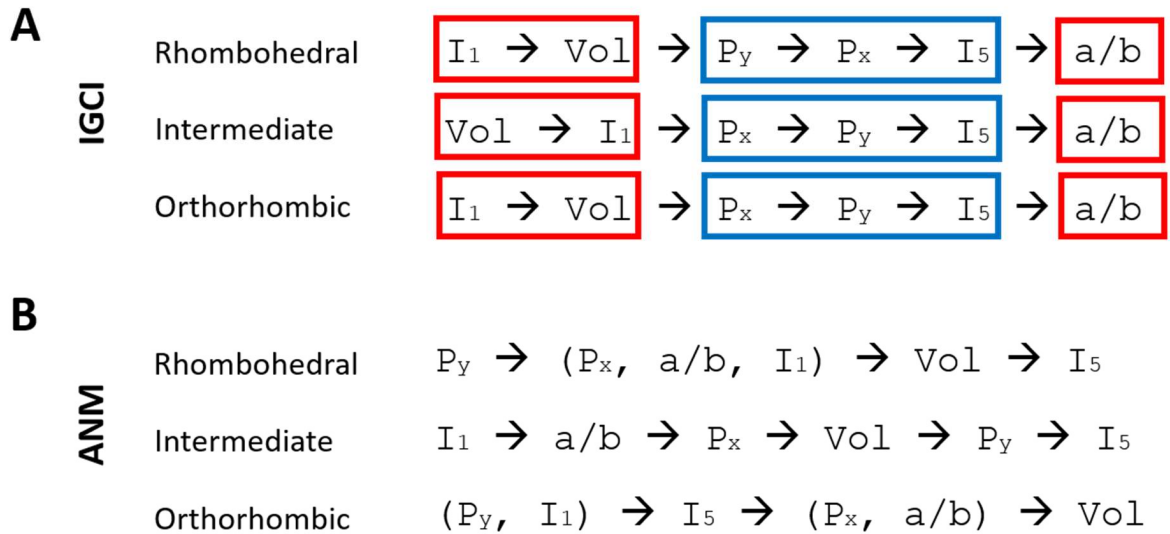

**Figure 6.** Causal chain analysis of the Sm-BFO system. Note that position of the descriptor in the chain suggest the likely cause and effect relationship, but still can be sensitive to the presence of confounders or observational bias.

The more direct way to visualize and compare the causal relationship is by arranging the observables as the causal chains as shown in Figure 6 for all three compositions. As seen in Figure 6, the dependence chain has significant overlaps between different compositions and analysis via IGCI and ANM. For IGCI, the chemical variables such as local composition and molar volume are clearly higher in the causal chain. In addition, the polarization components are arranged as ($P_y$, $P_x$) -> $I_5$ for all three compositions. This observation suggests that $I_5$ (differential contrast between A and B site cation intensity) is related to the physical distortion rather than chemical composition. Finally, tetragonality is the weakest variable for all the three phases and is ranked below the chemical and polarization components. While we are hesitant to drive the definitive conclusions from this analysis in the lack of large body of comparative studies, we note that this behavior generally comports to that expected from physics of material (except for $I_5$ variable)

For the ANM model, the analysis is less straightforward. Here, for the intermediate and orthorhombic phases, the chemical composition is identified as a more casually significant variable. This is in the agreement with the IGCI results. On the other hand, for the ferroelectric phase, one of the polarization components has the higher rank affecting molar volume tetragonality, etc. We note that this behavior is likely to be due to the nature of the ANM criterion, relying on the regression between the variable pairs. Given the fundamental difference between the ferroelectric and non-ferroelectric phase (presence of domains), this affects regression results and necessitates transition to Bayesian estimators.

To summarize, we have implemented pairwise causal analysis of the atomic scale structural, chemical, and polarization phenomena in the Sm-doped $BiFeO_3$ using scanning transmission electron microscopy data as descriptors. The causal sieve approach is implemented using the IGCI and ANM to establish the pairwise causal relationships between observables. The results can be represented as an ordered array of causal importance. For Sm-BFO compositions series studied here it is generally found via IGCI that the chemical effects including local composition and molar volume are higher on the causal chain and are not affected by polarization. The polarization effects are secondary, and differential chemical contrast and tetragonality are the weakest. The ANM analysis results are more difficult to interpret; here we argue that functional relationship between the variables are fundamentally different in dissimilar phases and therefore GP interpolation approach produces fundamentally different responses. This behavior will be explored in the future.

Overall, we note that optimization and discovery of new materials as well as understanding of fundamental physical mechanisms can be significantly accelerated if the causality of corresponding mechanisms can be understood, allowing to explore counterfactuals and interventions and avoiding correlative but incorrect conclusions. The fundamental physics offers a large set of knowledge on functional relationship between the materials parameters; however, real material systems often are characterized by only partially known physics or presence of non-equilibrated non-ergodic processes. We expect that in these cases causal analysis can provide the knowledge of cause and effect relationships necessary for materials optimization, design, and especially discovery.

**Acknowledgements:**

This effort (electron microscopy, feature extraction) is based upon work supported by the U.S. Department of Energy (DOE), Office of Science, Basic Energy Sciences (BES), Materials Sciences and Engineering Division (S.V.K., C.N.) and was performed and partially supported (M.Z., R.K.V.) at the Oak Ridge National Laboratory's Center for Nanophase Materials Sciences (CNMS), a U.S. Department of Energy, Office of Science User Facility. The work at the University of Maryland was supported in part by the National Institute of Standards and Technology Cooperative Agreement 70NANB17H301 and the Center for Spintronic Materials in Advanced infoRmation Technologies (SMART) one of centers in nCORE, a Semiconductor Research Corporation (SRC) program sponsored by NSF and NIST. A.N.M. work was partially supported by the European Union's Horizon 2020 research and innovation program under the Marie Skłodowska-Curie (grant agreement No 778070). 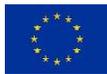 The authors express deepest gratitude for Prof. Judea Pearl (UCLA) and Dr. Vint Cerf (Google) for introduction in the field of causal machine learning and productive discussion.

**Materials:**

The combinatorial library of Sm – doped $BiFeO_3$ and the $SrRuO_3$ layer were both fabricated through pulsed laser deposition (PLD). Specifically, after reaching the base pressure (~ $2.0 \times 10^{-8}$ Torr) of the deposition chamber, the $SrTiO_3$ (001) substrate was heated up to 600 °C, and an oxygen flow was introduced to the chamber to maintain a desired deposition pressure (~ 100 mTorr). A laser energy density of ~ 0.8 $J/cm^2$ and an ablation frequency of 20 Hz were adopted for the deposition of the films. During the deposition of the $Bi_{1-x}Sm_xFeO_3$ layer, a $BiFeO_3$ target and a $SmFeO_3$ target were alternatively ablated, and a shadow mask was controlled to move accordingly to obtain a uniform composition gradient across the substrate.

**Methods:**

TEM samples were prepared by FIB liftout and local low energy Ar ion milling, down to 0.5 keV, in a Fischione NanoMill. STEM was performed at 200kV on a NION UltraSTEM. The three compositions were imaged consecutively to help maintain consistent imaging/microscope conditions. A correction algorithm was applied to correct for slow-scan axis scanning aberrations by reconstruction from two orthogonal source images according to [72]. All three datasets were defined with the [100] pseudocubic *a*-vectors along the thin film in-plane axis and the *b*-vector along the film growth axis. The atomic column positions ($A_{1,xy}$, $A_{2,xy}$, $A_{3,xy}$, $A_{4,xy}$, $B_{1,xy}$ inputs for polar displacement maps in Figure 1 and descriptors *a*, *b*, *a*Δ, *b*Δ, θ, Vol, & *P*) were determined by simultaneous 2D Guassian fits of local 5-atom perovskite unit cells. Atomic column HAADF intensity ($A_1$, $A_2$, $A_3$, $A_4$, and $B_1$ inputs for I1-I5 descriptors) was measured as the local Gaussian weighted 9-pixel intensity centered at the atom fit position. The source datasets were deliberately misaligned several degrees from the scan axes to aid the identification of residual scanning artifacts. Display images (Figure 1) & vector coordinates (*a*, *b*, *a*Δ, *b*Δ, & *P*) were subsequently rotated to align the mean *a*-vector to the horizontal axis. Calculation for the descriptors was performed according to Table 1. Unit cell grid maps for selected descriptors are shown in Figure 2 and in totality in the Supplemental Materials Figs 1-3) along with corresponding plotting information.

The causal data analysis is available in a form of executable Google Colab notebook at https://colab.research.google.com/github/ziatdinovmax/Notebooks-for-papers/blob/master/ferroics-causal-analysis.ipynb